\documentclass{jpp}
\usepackage{graphicx}
\usepackage{bm}
\usepackage[colorlinks,citecolor=blue,linkcolor=blue,urlcolor=blue]{hyperref}

\shorttitle{Simulations of relativistic reconnection with syn+IC cooling}
\shortauthor{K. Nalewajko, Y. Yuan and M. Chru{\'s}li{\'n}ska}

\title{Kinetic simulations of relativistic magnetic reconnection with synchrotron and inverse Compton cooling}

\author{Krzysztof Nalewajko\aff{1}
  \corresp{\email{knalew@camk.edu.pl}},
  Yajie Yuan\aff{2}
 \and Martyna Chru{\'s}li{\'n}ska\aff{3,4}}

\affiliation{
\aff{1}Nicolaus Copernicus Astronomical Center, Polish Academy of Sciences, ul. Bartycka 18, 00-716 Warsaw, Poland
\aff{2}Department of Astrophysical Sciences, Princeton University, Princeton, New Jersey, 08544, USA
\aff{3}Department of Astrophysics/IMAPP, Radboud University Nijmegen, P.O. Box 9010, 6500 GL Nijmegen, The Netherlands
\aff{4}Astronomical Observatory, University of Warsaw, al. Ujazdowskie 4, 00-478 Warsaw, Poland
}

\begin{document}

\maketitle

\begin{abstract}
First results are presented from kinetic numerical simulations of relativistic collisionless magnetic reconnection in pair plasma that include radiation reaction from both synchrotron and inverse Compton (IC) processes,
{motivated by non-thermal high-energy astrophysical sources, including in particular blazars.}
These simulations are initiated from a configuration known as ``ABC fields'' that evolves due to coalescence instability and generates thin current layers in its linear phase.
Global radiative efficiencies, instability growth rates, time-dependent radiation spectra, lightcurves, variability statistics and the structure of current layers are investigated for a broad range of initial parameters.
We find that the IC radiative signatures are generally similar to the synchrotron signatures.
The luminosity ratio of IC to synchrotron spectral components, the Compton dominance, can be modified by more than one order of magnitude with respect to its nominal value.
For very short cooling lengths, we find evidence for modification of the temperature profile across the current layers, no systematic compression of plasma density, and very consistent profiles of $\bm{E}\cdot\bm{B}$.
We decompose the profiles of $\bm{E}\cdot\bm{B}$ with the use of the Vlasov momentum equation, demonstrating a contribution from radiation reaction at the thickness scale consistent with the temperature profile.
\end{abstract}

\section{Introduction}

Many extreme astrophysical environments, especially those associated with black holes (active galactic nuclei, especially the relativistically beamed blazars; stellar X-ray binaries, especially microquasars) and neutron stars (magnetars, pulsars and their associated nebulae), as well as gamma-ray bursts, are efficient emitters of gamma-ray radiation. Their broad-band, power-law radiation spectra clearly indicate non-thermal, collisionless plasma environments that enable efficient particle acceleration to ultra-relativistic energies. In addition, many of these sources show powerful gamma-ray flaring activities with short variability time scales and high radiation efficiencies \citep[e.g.,][]{buehler_gamma-ray_2012,ackermann_minute-timescale_2016}, suggesting that the particle acceleration process {proceeds rapidly} and is likely a consequence of electromagnetic dissipation in a highly magnetised outflow from the central engine \citep[e.g.,][]{blandford_magnetoluminescence_2017}. 

{One of the most promising} dissipation and particle acceleration mechanisms is relativistic magnetic reconnection. Reconnection becomes relativistic when the magnetic energy density dominates the plasma rest-mass energy density, which may well be the case in the aforementioned astroplasma environments. Recently, significant progress has been made in understanding the basics of relativistic reconnection, with increasingly sophisticated fully kinetic numerical simulations employing the particle-in-cell (PIC) algorithm \citep[e.g.,][]{sironi_relativistic_2014,guo_formation_2014,werner_extent_2016,kagan_beaming_2016,sironi_plasmoids_2016}.

{Kinetic numerical simulations of relativistic magnetic reconnection are mostly local, starting from an initial configuration involving highly symmetric predefined Harris-type current layers.
In the context of global astrophysical systems like pulsar magnetospheres or relativistic jets of active galaxies, reconnection can be expected to be an intermittent process induced by dynamical collisions of misaligned magnetic domains or global current-driven instabilities \citep{mizuno_three-dimensional_2009,Lyutikov2018}.
Realistic (3D) global kinetic simulations of pulsar magnetospheres \citep{Philippov2015} or relativistic jets \citep{Nishikawa2016} are computationally very challenging.
However, intermittent reconnection with dynamical formation of kinetically thin current layers can now be investigated with a novel local magnetostatic configuration referred to as ``ABC fields'' (see Section \ref{sec_setup}).}

It has also been {possible}
to investigate the effects of radiation reaction to obtain self-consistent radiative signatures from particles accelerated by relativistic magnetic reconnection. The primary radiative mechanism investigated so far is synchrotron radiation with direct application to the problem of rapid gamma-ray flares from the Crab Nebula \citep[e.g.,][]{cerutti_simulations_2013,cerutti_three-dimensional_2014}, and synchro-curvature radiation with
application to the gamma-ray emission of pulsars \citep[e.g.,][]{cerutti_modelling_2016}.

In other cases, especially in blazars and microquasars, inverse Compton radiation can be equally important, or even dominate over the synchrotron mechanism.
Since blazars explain the bulk of cosmic gamma-ray radiation in the GeV and TeV bands \citep{Ajello2015}, this radiation is produced mainly in the inverse Compton process.
{The simplest radiative models of blazars assume that the observed synchrotron and IC spectral components are produced by the same population of energetic electrons.}
For a given blazar, the ratio of IC to synchrotron luminosities known as the Compton dominance ${\rm CD} = L_{\rm IC} / L_{\rm syn}$ is
{an indirect}
probe of plasma magnetisation in relativistic jets.
{If the IC scattering proceeds in the Thomson regime,
one can find that ${\rm CD} \sim U_{\rm rad}' / U_{\rm B}'$, where $U_{\rm rad}'$ is the energy density of soft radiation fields (external or synchrotron) and $U_{\rm B}'$ is the magnetic energy density, both evaluated in the jet co-moving frame}
\citep{Janiak2015}.

In this work, we present the first results from 2D and 3D PIC simulations of relativistic magnetic reconnection {initiated from the ``ABC fields''}, taking into account radiation reaction from both synchrotron and IC processes.\footnote{A recent study by \cite{Werner2018_IC} concerns the effect of IC radiation reaction without synchrotron cooling on particle energy distribution in highly relativistic reconnection initiated from the standard Harris layers, motivated primarily by accretion disk coronae.}
{The IC radiation is calculated as upscattering of fixed monoenergetic uniform isotropic radiation field, which corresponds to the ERC mechanism (Appendix \ref{sec_blazars}) operating in relativistic jets of the most luminous blazars.}
We investigate the spectral and temporal distributions of simulated radiative signatures of particle acceleration in dynamical current layers that form between highly magnetised coalescing magnetic domains. We also study the value of Compton dominance in relation to the adopted initial ratio of radiative and magnetic energy densities, as well as the perpendicular profiles of the current layers.

{Details of our numerical configuration are provided in Section \ref{sec_setup}.
Section \ref{sec_tot_ene} presents the results on the total energy transformations and radiative efficiency.
Section \ref{sec_rad_spe} describes the spectral distribution of simulated synchrotron and IC emission.
Section \ref{sec_lc} presents the results on time variability of simulated radiation signals.
In Section \ref{sec_prof} we investigate the effect of strong radiative cooling on the perpendicular structure of current layers.
Our results are discussed in Section \ref{sec_disc}.
In Appendix \ref{sec_blazars}, we provide a short summary of the observational properties of blazars and the characteristic physical parameters of their relativistic jets.}

\section{Numerical setup}
\label{sec_setup}

We perform particle-in-cell numerical simulations of periodic plasma configurations referred to as \emph{ABC fields} \citep[from Arnold-Beltrami-Childress; see][]{Dombre1986,east_spontaneous_2015,lyutikov_particle_2017}
using a modified version of the {\tt Zeltron} code \citep{cerutti_simulations_2013}.
In three dimensions, the magnetic fields are defined as:
\begin{eqnarray}
B_x(x,y,z) &=& B_0\left[\sin(\alpha z)+\cos(\alpha y)\right]\,,
\nonumber
\\
B_y(x,y,z) &=& B_0\left[\sin(\alpha x)+\cos(\alpha z)\right]\,,
\\
B_z(x,y,z) &=& B_0\left[\sin(\alpha y)+\cos(\alpha x)\right]\,,
\nonumber
\end{eqnarray}
where $\alpha = 2\pi k/L$ and $L$ is the linear size of the simulation domain: $x,y,z\in[0:L]$. The case of $k = 1$ corresponds to the lowest-energy stable Taylor state. The case of $k = 2$ is the lowest-energy unstable configuration.
In 2D there exists a smaller unstable configuration obtained by rotating the coordinate system \citep{nalewajko_kinetic_2016,yuan_kinetic_2016}.
\begin{eqnarray}
\label{eqn_ic}
B_x(x,y) &=& B_0\left[\sin(\alpha(x+y))+\sin(\alpha(x-y))\right]/\sqrt{2}\,,
\nonumber
\\
B_y(x,y) &=& B_0\left[\sin(\alpha(x-y))-\sin(\alpha(x+y))\right]/\sqrt{2}\,,
\\
B_z(x,y) &=& B_0\left[\cos(\alpha(x+y))-\cos(\alpha(x-y))\right]\,,
\nonumber
\end{eqnarray}
We will refer to this configuration as ``diagonal ABC fields''.
{We note that this initial magnetic field distribution is characterised by $\max(B) = 2B_0$ and $\left<B^2\right> = 2B_0^2$.}

In order to obtain a rough equilibrium, current density $\bm{j} = (kc/L)\bm{B}$ is provided by a population of relativistic {\rm electrons and positrons} characterised by uniform total number density $n = 3\sqrt{2}kB_0/(2e\tilde{a}_1L)$,
{by the Maxwell-J\"{u}ttner energy distribution with relativistic temperature $\Theta_{\rm e} = k_{\rm B}T_{\rm e}/(m_{\rm e}c^2)$}, and by the dipole moment of the local angular distribution of particle momenta $a_1 = (B/B_0)\tilde{a}_1$, and $\tilde{a}_1 = B_0/\max(B) = 0.5$ is a constant.
{As we work here in the limit of highly relativistic electron temperatures $\Theta_e \gg 1$, we use the following statistics of the Maxwell-J\"{u}ttner distribution $\left<\gamma\right> \simeq 3\Theta_e$ and $\left<\gamma^2\right> \simeq 12\Theta_e^2$.
The corresponding mean hot magnetisation value is given by $\left<\sigma_{\rm hot}\right> = \left<B^2\right>/(4\pi w)$, where $w \simeq 4\Theta_enm_{\rm e}c^2$ is the ultra-relativistic specific enthalpy:
\begin{equation}
\left<\sigma_{\rm hot}\right> \simeq \frac{\tilde{a}_1}{12\sqrt{2}\pi k}\left(\frac{L}{\rho_0}\right)
\,,
\end{equation}
The characteristic property of ABC fields is that magnetisation scales linearly with the scale separation between the magnetic field coherence scale (of order $L$) and the kinetic gyration scale ($\rho_0$) \citep{nalewajko_kinetic_2016}. This is because a minimum particle number density is required in order to realise smoothly distributed current density.
}

Radiation reaction from synchrotron and inverse Compton processes was included in advancing the particle momenta. For the synchrotron process, local values of magnetic and electric fields $\bm{B},\bm{E}$ are used
{for an electron with velocity $\bm{v} = \bm\beta c$, Lorentz factor $\gamma = (1-\beta^2)^{-1/2}$ and dimensionless 4-velocity $\bm{u} = \gamma\bm\beta$
\citep{cerutti_simulations_2013}:
\begin{eqnarray}
\frac{\partial \bm{u}}{\partial t} &=& -\frac{P_{\rm syn}\bm{u}}{\gamma m_ec^2}\,,
\qquad
P_{\rm syn} = \frac{\sigma_{\rm T}c}{4\pi}\left[(\gamma\bm{E}+\bm{u}\times\bm{B})^2-(\bm{E}\cdot\bm{u})^2\right]\,.
\end{eqnarray}
}
For the inverse Compton process, we assume a uniform isotropic external radiation field parametrized by energy density $U_{\rm ext}$ and photon energy $E_{\rm ext}$.
In order to account for the Klein-Nishina cross section in the Zeltron code, we have updated the radiation reaction formula of \cite{cerutti_simulations_2013} \citep{Jones1968}:
\begin{eqnarray}
\frac{\partial \bm{u}}{\partial t} &=& -\frac{P_{\rm IC}\bm{u}}{\gamma m_ec^2}\,,
\qquad
P_{\rm IC} = \frac{4}{3}\sigma_T c U_{\rm ext} u^2 f_{\rm KN}(b)\,,
\qquad
b = \frac{4\gamma E_{\rm ext}}{m_{\rm e}c^2}\,,
\\
f_{\rm KN}(b) &=& \frac{9}{b^3}\left[\left(\frac{b}{2} + 6 + \frac{6}{b}\right)\ln(1+b) - \frac{1}{(1+b)^2}\left(\frac{11}{12}b^3 + 6b^2 + 9b + 4\right)\right.
\\
&&\left.- 2 + 2{\rm Li}_2(-b)\right]\,.
\quad
\nonumber
\end{eqnarray}

{The synchrotron radiation spectra are calculated by summing spectral contributions from all individual macroparticles (electrons and positrons) of velocity $\bm{v}$ and Lorentz factor $\gamma$ \citep{Blumenthal_Gould1970}:
\begin{eqnarray}
L_{\rm syn}(\nu) &=& \frac{\sqrt{3}e^2}{c}\sum_{\rm e^+e^-}N_{e,1}F(\xi)\Omega_{\rm syn}\,,
\\
F(\xi) &=& \xi\int_{\xi}^{\infty}K_{5/3}(x){\rm d}x\,,
\\
\xi &=& \frac{4\pi\nu}{3\gamma^2\Omega_{\rm syn}}\,,
\qquad
\Omega_{\rm syn} = \frac{e}{m_ec}\left|(\bm{E} + \bm{n}\times\bm{B})\times\bm{n}\right|\,,
\end{eqnarray}
where $\bm{n} = \bm{v}/|\bm{v}|$ is the unit vector along the particle velocity, $N_{e,1}$ is the number of electrons represented by individual macroparticle, and $K_a$ is the modified Bessel function of the second kind.
Because we consider ultra-relativistic particles with $\Theta_e \gg 1$, our approach is self-consistent for $\xi = \omega/\omega_{\rm c} > (\omega_c \Delta t)^{-1} = (5\Theta_{\rm e}^3)^{-1}$, where $\omega_{\rm c} = (3/2)\gamma^2\Omega_{\rm syn} \simeq 18\Theta_{\rm e}^3(c/\rho_0)$ and $\Delta t = 0.99 \Delta x/(\sqrt{2}c)$ is the simulation time step, and hence there is no need for a more general Fourier transformation method of \cite{Hededal2005}.

The IC radiation spectra are calculated in an analogous way, using the general Klein-Nishina kernel \citep{Blumenthal_Gould1970}:
\begin{eqnarray}
L_{\rm IC}(\nu) &=& \frac{12h\sigma_{\rm T}U_{\rm ext}}{m_ec}\sum_{\rm e^+e^-}N_{e,1}\frac{\gamma\xi}{b^2}K(\xi,b)\,,
\\
K(\xi,b) &=& 1 + q - 2q^2 + 2q\log{q} +  q(1-q)\frac{b\xi}{2}\,,
\\
q &=& \frac{\xi}{b(1-\xi)} < 1\,,
\qquad
\xi = \frac{h\nu}{\gamma m_ec^2} < 1\,,
\qquad
b = \frac{4\gamma E_{\rm ext}}{m_ec^2}\,.
\end{eqnarray}
}

We performed two series of 2D simulations initiated from diagonal ABC fields with $k = 1$ on numerical grid of size $N_x = N_y = 2048$ at resolution $\Delta x^i = \rho_0/2.56$, where $\rho_0 = \Theta_{\rm e} m_{\rm e} c^2/(e B_0)$ is the nominal gyroradius, {and with 128 particles per cell}. In all cases, the energy density of external radiation fields was set at {$U_{\rm ext} = \left<B^2\right>/8\pi = B_0^2/4\pi = 2U_0$, where $U_0 = B_0^2/8\pi$ is the nominal magnetic energy density.
The domain size of $L = 800\rho_0$ corresponds to the mean hot magnetisation value of $\left<\sigma_{\rm hot}\right> \simeq 7.5$.}
Parameters of all PIC simulations presented in this work are compared in Table \ref{tab:sims}.

\begin{table}
\caption{List of PIC simulations described in this work. {$L/\rho_0$ is the physical size of the simulation domain in units of nominal gyroradius $\rho_0 = \Theta_{\rm e}m_{\rm e}c^2/(eB_0)$, $B_0$ is the nominal magnetic field strength in units of Gauss, $\Theta_{\rm e} = k_{\rm B}T_{\rm e}/(m_{\rm e}c^2)$ is the dimensionless relativistic temperature of the initial particle energy distribution, $E_{\rm ext}$ is the photon energy of soft radiation in units of eV, $l_{\rm cool}$ is the nominal cooling length defined in Eq. (\ref{eq:lcool}), and $\tau_{\rm E}$ is the exponential growth rate of total electric energy in units of light-crossing time scale $L/c$.}}
\label{tab:sims}
\centering
\vskip 1em
\begin{tabular}{lcccccc}
name & $L/\rho_0$ & $B_0$[G] & $\Theta_{\rm e}$ & $E_{\rm ext}$[eV] & $l_{\rm cool}/\rho_0$ & $c\tau_{\rm E}/L$ \\
\hline
2Da\_T1e5   & 800 &              1 &         $10^5$ & 0.1 & $8.5\times 10^4$ & 0.182 \\
2Da\_T3e5   & 800 &              1 & $3\times 10^5$ & 0.1 & $9.5\times 10^3$ & 0.180 \\
2Da\_T1e6   & 800 &              1 &         $10^6$ & 0.1 & 850 & 0.171 \\
2Da\_T3e6   & 800 &              1 & $3\times 10^6$ & 0.1 & 95 & 0.158 \\
2Da\_T1e7   & 800 &              1 &         $10^7$ & 0.1 & 8.5 & 0.144 \\
\hline
2Db\_fc\_kn & 800 & $2\times 10^4$ & $10^4$ &  40 & 430 & 0.173 \\
2Db\_fc\_th & 800 & $2\times 10^4$ & $10^4$ & 0.4 & 430 & 0.164 \\
2Db\_sc\_kn & 800 & $2\times 10^2$ & $10^4$ &  40 & $4.3\times 10^4$ & 0.182 \\
2Db\_sc\_th & 800 & $2\times 10^2$ & $10^4$ & 0.4 & $4.3\times 10^4$ & 0.181 \\
\hline
3D          & 900 &              1 & $10^6$ & 0.01 & 850 & 0.189 \\
\end{tabular}
\vskip 1em
\end{table}

The first series, denoted as `2Da', was performed for the same value of $B_0 = 1\;{\rm G}$ and for different values of initial particle temperature: $\Theta_{\rm e} = 10^5,3\times 10^5,10^6,3\times 10^6,10^7$.
Increasing the particle temperature leads to increasing the efficiency of radiative cooling, and hence a shorter cooling length.
The nominal cooling length {with IC cooling in the Thomson limit} is given by:
\begin{equation}
\label{eq:lcool}
l_{\rm cool} = c\tau_{\rm cool} = \frac{\left<\gamma\right>}{\left<|{\rm d}\gamma/c{\rm d}t|\right>} = 
\frac{\left<\gamma\right>}{\left<\gamma^2\right>}\frac{3m_{\rm e}c^2}{4\sigma_TU_{\rm cool}}
\simeq
\frac{(3\pi/8)e}{\sigma_T\Theta_e^2B_0}\rho_0
\end{equation}
{where $U_{\rm cool} = \left<U_B\right> + U_{\rm ext} = 4U_0$ is the effective cooling energy density.}
The values of $l_{\rm cool}/\rho_0$ are reported in Table \ref{tab:sims}.

The second series, denoted as `2Db', was performed for the same value of initial particle temperature: $\Theta_{\rm e} = 10^4$, and for different values of $B_0$ and $E_{\rm ext}$.
Two of these simulations were performed in the fast-cooling (fc) regime by setting $B_0 = 2\times 10^4\;{\rm G}$, and two simulations were performed in the slow-cooling regime (sc) by setting $B_0 = 2\times 10^2\;{\rm G}$.
{We note that such high magnetic field strengths have been suggested in the context of the most extreme GeV gamma-ray variability of blazar 3C~279 observed by Fermi/LAT on suborbital time scales of a few minutes \citep{ackermann_minute-timescale_2016}.}
Furthermore, two simulations were performed with Compton scattering in the Klein-Nishina (kn) regime by setting $E_{\rm ext} = 40\;{\rm eV}$, and two simulations were performed with Compton scattering in the Thomson (th) regime by setting $E_{\rm ext} = 0.4\;{\rm eV}$.

All of our 2D simulations show the same basic behaviour as the ones reported in earlier works \citep{nalewajko_kinetic_2016,yuan_kinetic_2016}. Initial equilibrium evolves due to coalescence instability, which leads to the formation of dynamical current layers over a single light-crossing timescale $L/c$. This is followed by slowly damped non-linear global oscillations. Particles are accelerated most efficiently by electric fields present in linear current layers.

We are also reporting partial results from a single 3D simulation {for ``standard'' ABC fields with $k=2$ and $\tilde{a}_1 = 0.4$, that was performed with the same prescription for radiative losses.
This simulation was performed on numerical grid of size $N_x = N_y = N_z = 1152$ at resolution $\Delta x^i = \rho_0/1.28$ with 16 particles per cell.}
Detailed results of this simulation will be presented in another publication.

\section{Global energy transformations and radiative efficiency}
\label{sec_tot_ene}

\begin{figure}
  \includegraphics[width=\textwidth]{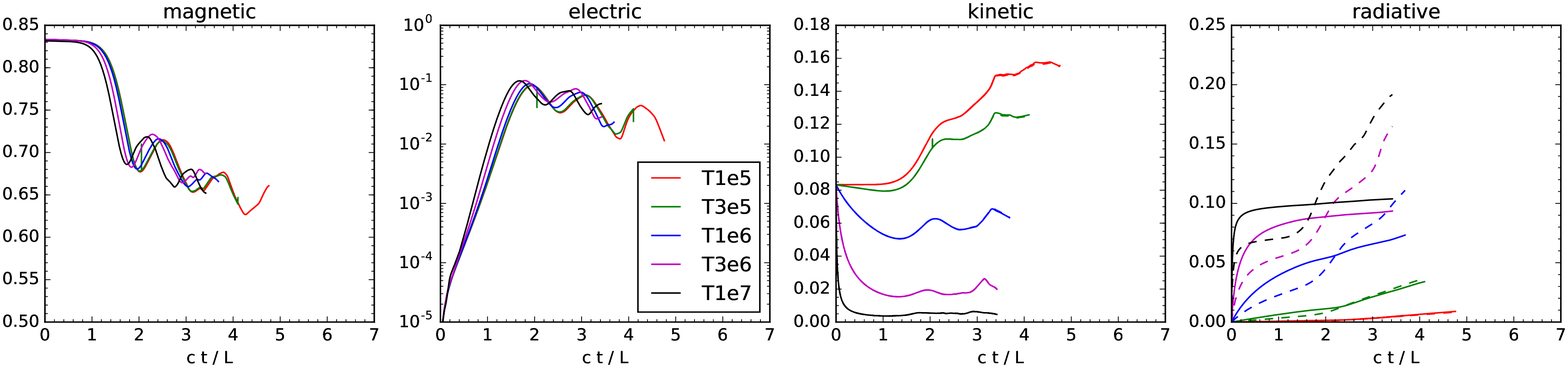}
  \includegraphics[width=\textwidth]{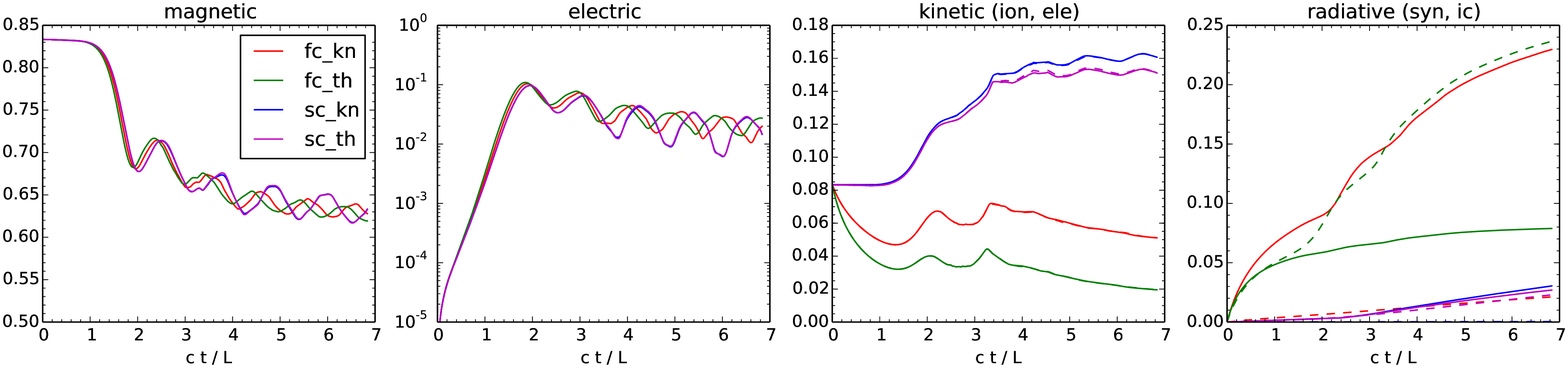}
  \includegraphics[width=\textwidth]{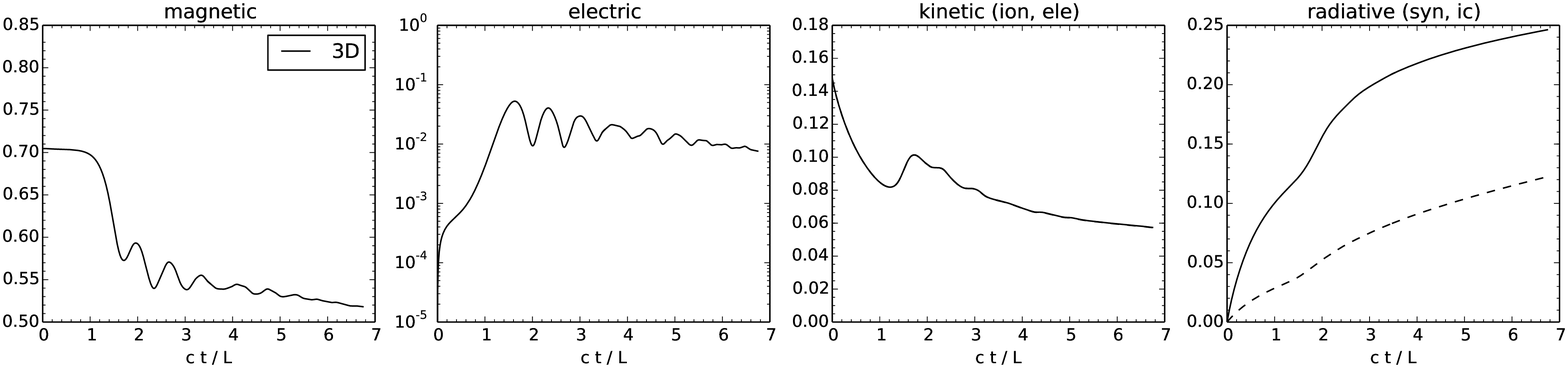}
  \caption{Components of the total energy as functions of simulation time, {normalised to the total energy including the total radiation losses}. Top row: the `2Da' series; middle row: the `2Db' series; bottom row: the 3D simulation. {In the third column, the solid lines show the contribution of positrons, and the dashed lines show the contribution of electrons. In the last column, the solid lines show integrated synchrotron energy losses, and the dashed lines show the integrated IC energy losses.}}
  \label{fig:tot_ene}
\end{figure}

Figure \ref{fig:tot_ene} compares the time evolutions of total energy components, which include the magnetic, electric, kinetic energies, as well as the global energy radiated away in the synchrotron and inverse Compton processes.

\subsection{Results of the 2D simulations}

In all our 2D simulations, the total energy of the system is dominated by magnetic energy, with the initial share of 83\%, decreasing to $\simeq 68\%$ by the moment of saturation of the linear instability ($ct/L \simeq 1.9$), followed by a further slow decrease to $\simeq 63\%$ by $ct/L \simeq 7$. Evolution of the magnetic energy fraction is very similar in all cases. The electric energy reaches a peak of $\simeq 10\%$ at the saturation point, with slightly higher values
in the fast-cooling regimes. Major differences are seen in the evolutions of energy shared by the particles.
In the `2Da' series, the initial fraction of $\simeq 16\%$ (including both electrons and positrons) increases to $\sim 30\%$ for $\Theta_{\rm e} = 10^5$, but it decreases down to $<1\%$ for $\Theta_{\rm e} = 10^7$.
In the `2Db' series, the kinetic energy fraction reaches between $\simeq (30-32)\%$ in the slow-cooling cases, and $<10\%$ in the fast-cooling cases.
These differences are reflected in the radiative efficiencies, which account for cumulative radiative energy losses of all particles. In the slow-cooling cases, the radiative efficiencies of our simulations are of order $\sim (3-5)\%$, counting both synchrotron and IC channels.

The most interesting result here is the radiative efficiency in the fast-cooling regime.
In the `2Da' series, by $ct/L = 3$, the synchrotron radiative efficiency reaches $10\%$ for $\Theta_{\rm e} = 10^7$, and the IC radiative efficiency becomes even higher, up to $\simeq 17\%$.
We also find that the IC radiative efficiency dominates the synchrotron radiative efficiency for $\Theta_{\rm e} \ge 10^6$, while they are comparable to each other for $\Theta_{\rm e} < 10^6$.
In the `2Db' series, by $ct/L = 7$, with IC cooling in the Klein-Nishina regime (fc\_kn), we achieved a synchrotron radiative efficiency of $\simeq 23\%$, while the IC radiative efficiency was only $\simeq 2\%$.
However, in the Thomson regime (fc\_th), the IC radiative efficiency of $\simeq 23\%$ dominates the synchrotron efficiency of $\simeq 8\%$, and this is despite these efficiencies being initially ($ct/L < 1$) comparable due to our choice of $U_{\rm ext}/U_{\rm B0}$.
While $U_{\rm ext}$ is constant, the magnetic energy density $U_{\rm B}$ decreases during the simulation by $\simeq 25\%$.
On the other hand, the ratio of cumulative radiative efficiencies is $f_{\rm IC}/f_{\rm syn} \sim 3$ (we also note that this ratio is less than unity in the sc\_th case),
which means that the local magnetic field component perpendicular to the velocities of emitting particles is systematically reduced. There can be two reasons for this: (a) that particles are radiating preferentially in regions of low magnetic field strength, (b) that particles are radiating preferentially along local magnetic field lines. Both possibilities are consistent with emission being produced primarily within the reconnection current layers and their outflowing nozzles \citep{yuan_kinetic_2016}.

We also measure growth timescales of the linear coalescence instability defined as the $e$-folding timescale $\tau$ of the total electric energy $E^2$, normalised to the light-crossing timescale $L/c$.
In the `2Da' series, we find systematic decrease of growth timescale with increasing cooling efficiency, from $c\tau/L \simeq 0.18$ for $\Theta_{\rm e} \le 3\times 10^5$ to $c\tau/L \simeq 0.145$ for $\Theta_{\rm e} = 10^7$. 
In the `2Db' series, the longest growth timescale $c\tau/L \simeq 0.165$ is measured in the `fc\_th' case, and the shortest growth timescales of $c\tau/L \simeq 0.18$ are measured in the `sc' cases.
We thus find a consistent picture that the instability growth timescale is enhanced by strong radiative cooling, which also results in removing the gas pressure and increasing effective magnetisation $\sigma$.
This is consistent with the previous result showing that the growth timescale decreases systematically with $\sigma$ and is longer than the force-free limit of $c\tau_{\rm ff}/L \simeq 0.13$ \citep{nalewajko_kinetic_2016}.

\subsection{Results of the 3D simulation}

Figure \ref{fig:tot_ene} shows that the initial state is characterised by lower magnetisation, with magnetic fields containing roughly $\sim 71\%$ of the total energy. Correspondingly, more energy ($\sim 29\%$; summing electrons and positrons) is contained in the particles. The overall cooling rate is very fast, so that before magnetic reconnection is able to compete with the radiative cooling, the kinetic energy content of particles decreases to $\sim 16\%$. The exponential growth timescale of electric energy is $c\tau/L \simeq 0.19$,
but it saturates at the level of $\sim 5\%$, just half of the 2D level.
However, noting that our 3D configuration is characterised by ABC wavelength shorter by factor $\sqrt{2}$ compared with the 2D configuration, the effective growth timescale of the 3D simulation would amount to $0.27$.
By $ct/L = 6.5$, the radiative efficiency of the synchrotron mechanism is $\simeq 24\%$, dominating the radiative efficiency of the IC mechanism of $\simeq 12\%$. The Klein-Nishina parameter $b \simeq 0.25$ is somewhat higher than in the case 2Db\_fc\_th ($b \simeq 0.1$). Although still technically in the Thomson regime, the difference in $b$ values can partly explain the difference between synchrotron-dominated radiative losses in 3D and the IC-dominated losses in 2D.

\begin{figure}
  \centering
  \includegraphics[width=\textwidth]{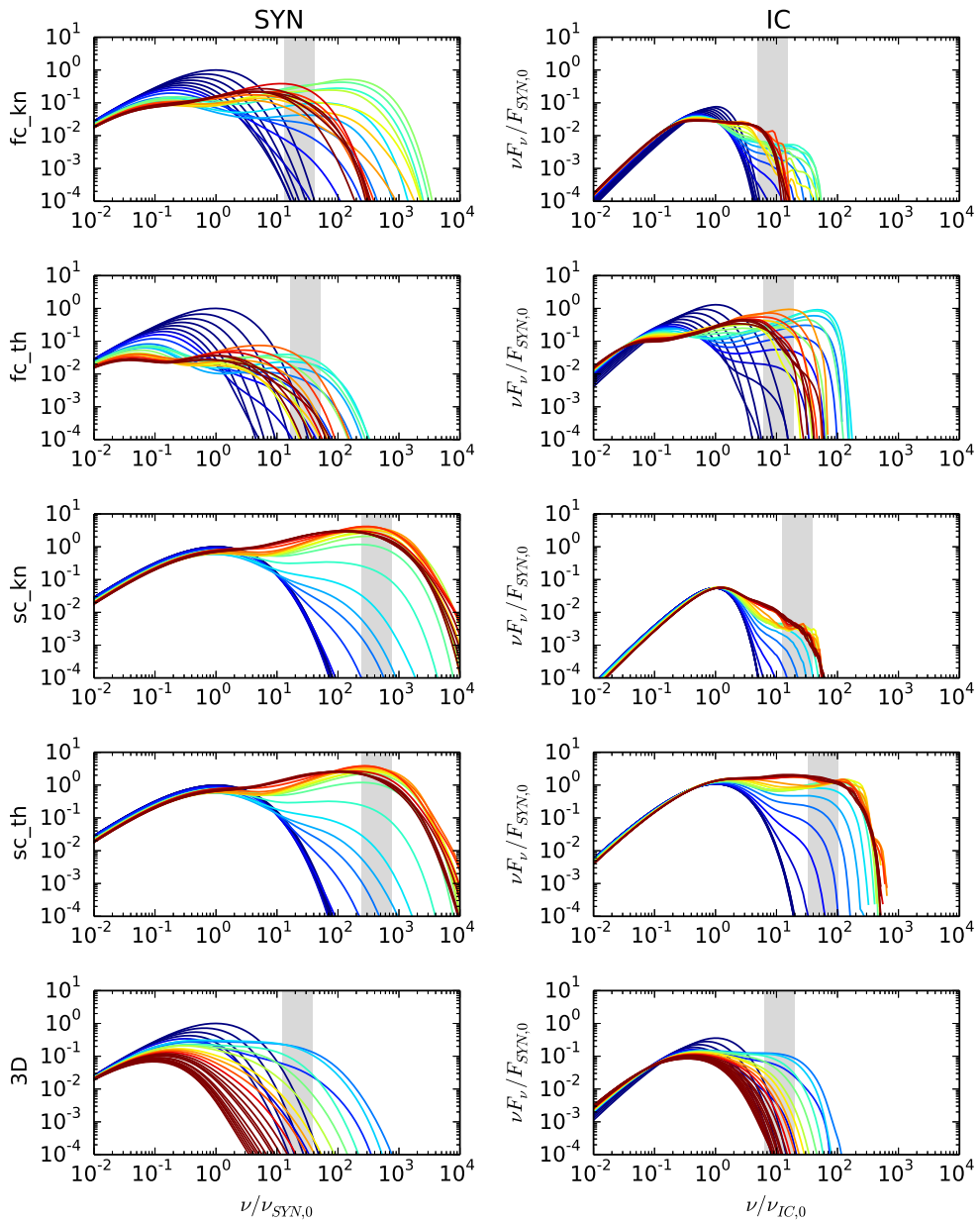}
  \caption{Time evolution of radiation spectra -- synchrotron (left panels) and IC (right panels) -- compared for simulations in the `2Db' series and for the 3D simulation. {The line colours indicate the simulation time progressing from deep blue ($ct/L \le 1$) to brown ($ct/L \ge 4$).} The spectra are normalised to the initial peak of the synchrotron component. The grey stripes indicate the frequency ranges from which lightcurves presented in Figure \ref{fig:lc} are extracted.}
  \label{fig:spe}
\end{figure}

\section{Spectral energy distribution of radiation}
\label{sec_rad_spe}

\begin{figure}
  \centering
  \includegraphics[width=\textwidth]{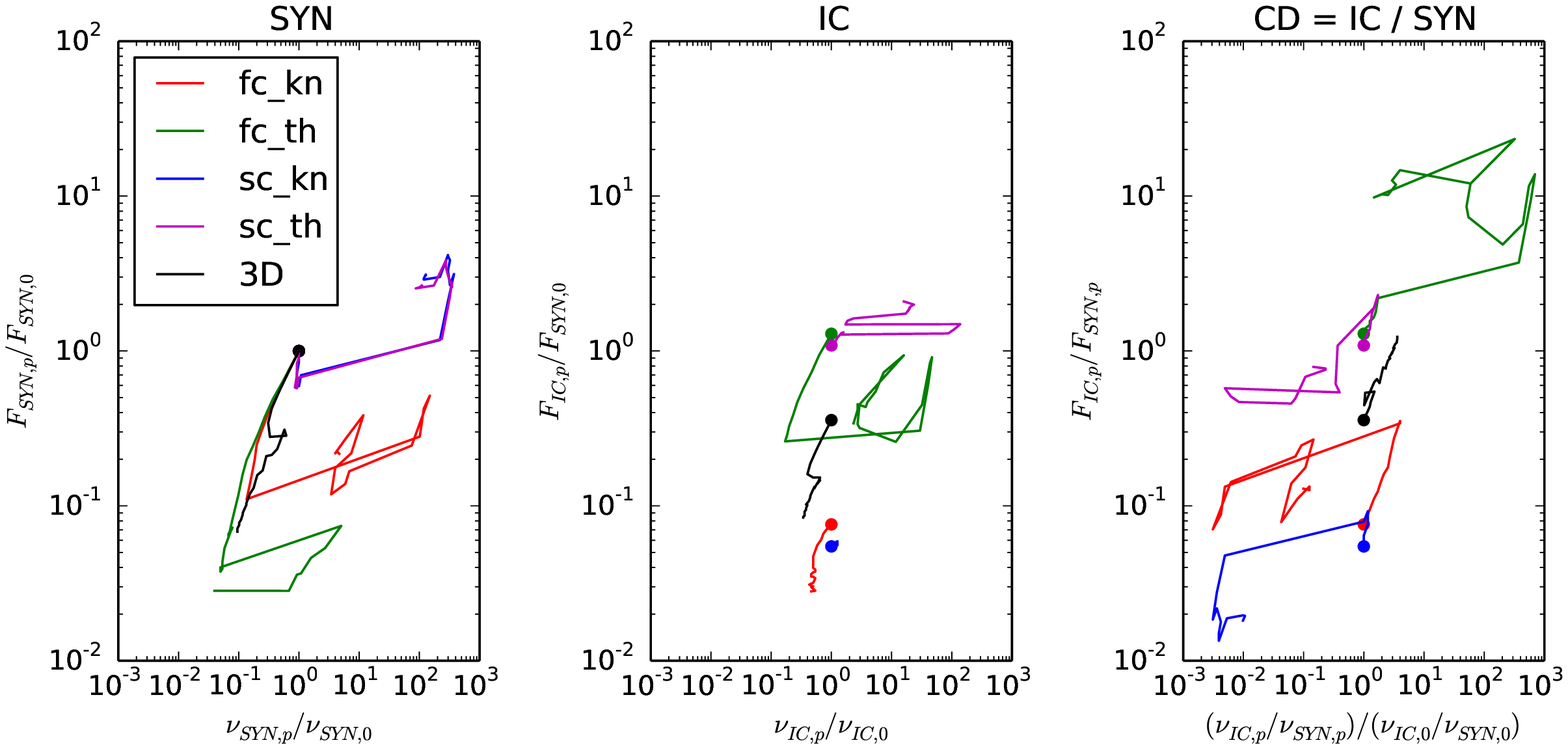}
  \caption{Time evolution of radiation SED peaks -- synchrotron (left panel), IC (middle panel) and Compton dominance (IC/SYN; right panel) -- compared for simulations in the `2Db' series and for the 3D simulation.}
  \label{fig:peaks}
\end{figure}

Figure \ref{fig:spe} shows the time evolution of the $\nu F_\nu$ radiation spectral energy distributions (SED), showing separately the synchrotron and IC components. All panels are scaled in the same way with respect to the initial synchrotron luminosity $F_{\rm SYN,0}$, and to the respective peak frequencies. Corresponding to these SEDs, Figure \ref{fig:peaks} compares the positions $(\nu_p,F_p)$ of respective SED peaks.

\subsection{Results of the 2Db simulations}

In the slow-cooling regime, we can see how a high-frequency synchrotron component develops due to energetic particles emerging at the sites of magnetic reconnection. This component has peak frequency of $\nu_{\rm syn,p} \sim 200 \nu_{\rm syn,0}$, and peak luminosity of $F_{\rm syn,p} \sim 3F_{\rm syn,0}$. This component does not depend on whether the IC cooling proceeds in the Thomson or Klein-Nishina regime. However, a major difference can be seen in the IC component: as can be expected, it is significantly suppressed in the `sc\_kn' case, and the SED remains dominated by low-energy electrons. In the `sc\_th' case, a high-frequency peak is obtained at $\nu_{\rm IC,p} \sim 100\nu_{\rm IC,0}$ and $F_{\rm IC,p} \sim F_{\rm IC,0}$. The Compton dominance in the `sc\_th' case is in the range ${\rm CD} \in [0.5:1]$, while it decreases to $\sim 0.02$ in the `kn' case. The peak frequency ratio is of order $\nu_{\rm IC,p}/\nu_{\rm syn,p} \sim 0.01(\nu_{\rm IC,0}/\nu_{\rm syn,0})$.

The situation is different in the fast-cooling regime, where strong radiative losses force a rapid evolution of the initial SEDs. Initially, the synchrotron luminosity decreases more strongly than the IC luminosity, and hence the Compton dominance increases. In the `fc\_th' case, this increase is by one order of magnitude, and we have ${\rm CD} \sim 10$ through the end of simulation. In the `fc\_kn' case, the increase is by factor 4, but it is reversed due to rebrightening of the synchrotron component. The latter case is also characterised by minor evolution of the IC component.

\subsection{Results of the 3D simulation}

Figure \ref{fig:spe} shows fairly similar time evolution of the synchrotron and IC SED components. Figure \ref{fig:peaks} demonstrates similar initial behaviours of the synchrotron peaks between 3D and 2Db\_fc\_th, and a notably stable location of the IC/SYN peak ratio with Compton dominance of the order of unity.

\section{Lightcurves}
\label{sec_lc}

\begin{figure}
  \centering
  \includegraphics[width=\textwidth]{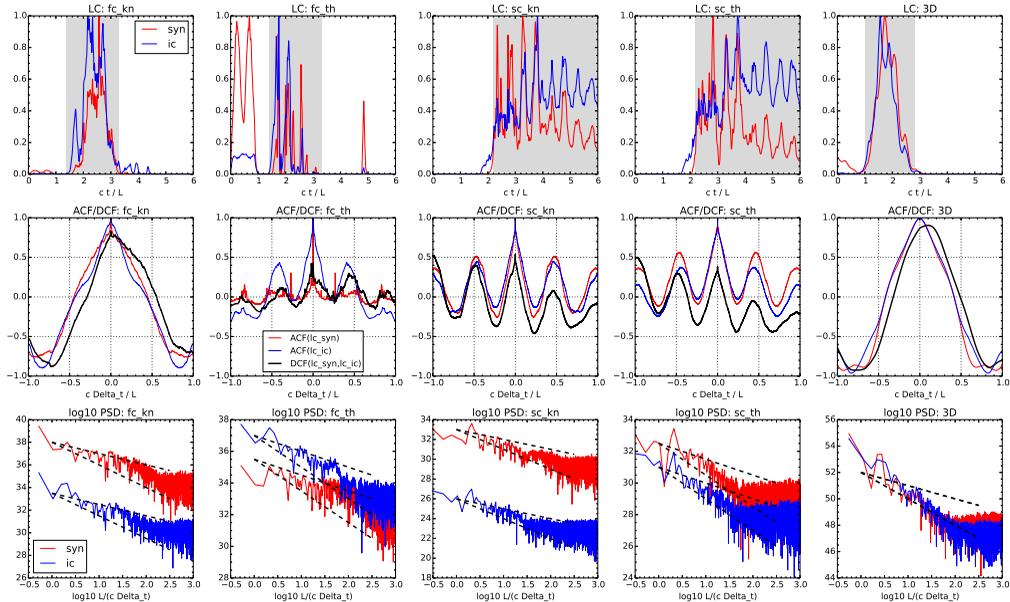}
  \caption{
  \emph{Top row:} Lightcurves calculated for simulations in the `2Db' series and for the 3D one in the frequency ranges indicated in Figure \ref{fig:spe}. These lightcurves are smoothed and normalised independently to their respective maximum values.
  \emph{Middle row:} Autocorrelation functions (ACF) and discrete correlation functions (DCF) calculated for the sections of lightcurves (without smoothing) indicated by shaded stripes in the respective top panels.
  \emph{Bottom row:} Power spectral densities (PSD) calculated for the sections of lightcurves (without smoothing nor normalisation) indicated by shaded stripes in the respective top panels. The dashed black lines indicate power-law indices $-1$ and $-2$ for reference.
  }
  \label{fig:lc}
\end{figure}

We calculate lightcurves of synchrotron and inverse Compton radiation for a specific observer and for several frequency bands, for all simulations in the `2Db' series and for the 3D one. Figure \ref{fig:spe} shows the frequency bands used for calculating the lightcurves presented in Figure \ref{fig:lc}. We focus our attention on the spectral bands dominated by contribution from particles accelerated by magnetic reconnection.

In Figure \ref{fig:lc}, we compare directly synchrotron and IC lightcurves normalised independently to unity, as would be observed simultaneously by the same observer. Major difference can be seen between the fast- and slow-cooling regimes.
In the former (including the 3D case), a brief flash lasting a little over a single light-crossing time $\Delta (ct/L) \sim 1$ would be seen.
In the `fc\_th' case, we also see a strong signal of synchrotron radiation for $ct/L < 1$ produced by the initial population of very hot particles.
In the slow-cooling regime, the light-curves are more extended in time, and they are expected by decay at a very slow rate.
A quasi-periodic pattern at the period of $\sim L/(2c)$ is apparent in both synchrotron and IC lightcurves. No major difference is seen between the Thomson and Klein-Nishina regimes of the IC process.

In Figure \ref{fig:lc}, we also present the temporal correlation analysis of the calculated lightcurves, using the method of Discrete Correlation Function (DCF; \citealt{EdelsonKrolik1988}).
{The DCF method takes two series of measurements $(t_{1,i},f_{1,i})$ and $(t_{2,i},f_{2,i})$, where time sampling can be irregular, bins all pairs of measurements according to time delay $\Delta t = t_{2,j} - t_{1,i}$, and evaluates the average discrete correlation $\left<\tilde{f}_{1,i}\tilde{f}_{2,j}\right>$ for each time delay bin using normalised measurements $\tilde{f}_i = (f_i - \left<f\right>)/\sigma_f$, where $\sigma_f$ is a dispersion of $f_i$. The result of DCF analysis is correlation coefficient as function of time delay ${\rm DCF}(\Delta t) \in [-1:1]$.}
We calculate DCF(SYN,IC) between the synchrotron and IC lightcurves, as well as Autocorrelation Functions ACF(X)=DCF(X,X) for either the synchrotron or IC lightcurves. The ACFs reveal characteristic variability timescales (measured between two zero points bracketing the main peak) of $\simeq 0.8(L/c)$ in the cases 2Db\_fc\_kn and 3D, and $\simeq 0.4(L/c)$ in the 2Db\_sc cases. The case 2Db\_fc\_th is more ambiguous, as the lightcurves in this case are barely resolved in time. No major differences are seen between the ACFs calculated for the synchrotron or IC lightcurves. With this insight from the ACFs, the DCF between synchrotron and IC lightcurves can tell us two things: the peak correlation coefficient, and the presence of any time lag between the two lightcurves. We find that the correlation coefficient for the cases with longer variability timescales (2Db\_fc\_kn and 3D) is $\ge 0.75$, while in the cases with shorter variability timescales (2Db\_sc) it is $\le 0.5$. A minor time lag of $+0.1(L/c)$ (synchrotron lagging the IC) is indicated only for the 3D case, however, it is very unlikely to be statistically significant, as ${\rm DCF}(\Delta t = 0) = 0.87$.

Also in Figure \ref{fig:lc}, we present the variability statistics in the form of Power Spectral Density (PSD) calculated separately for the synchrotron and the IC lightcurves. We demonstrate that in all `2Db' cases, the PSD is consistent with a power-law with index between $-2$ (red noise) and $-1$ (pink noise). However, in the 3D case, the PSD is significantly steeper, with the index close to $-3$.

\section{Perpendicular profiles of the current layers}
\label{sec_prof}

\begin{figure}
  \centering
  \includegraphics[width=\textwidth]{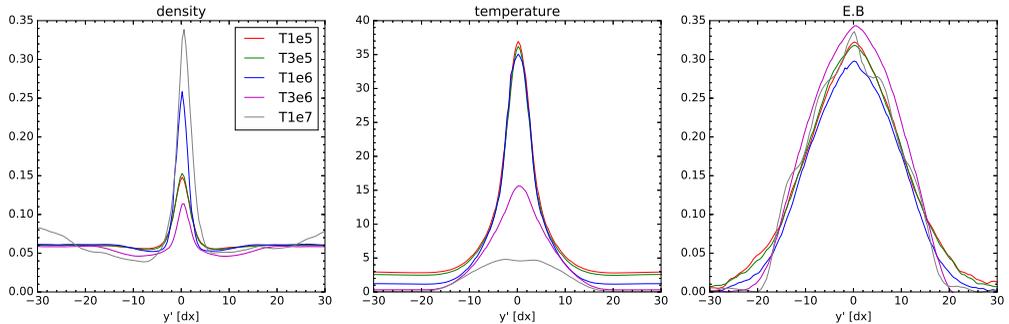}
  \caption{Perpendicular profiles of current layers for simulations in the `2Da' series with different particle temperatures $\Theta$, and hence different radiative cooling time scales. The panels present: the particle number density $n$ (left), the mean particle energy $\left<\gamma\right>$ (middle), and the $\bm{E}\cdot\bm{B}$ (right). For each $\Theta$ value, the profiles were extracted at the simulation time that maximises the amplitude of $\bm{E}\cdot\bm{B}$.}
  \label{fig:prof_perp}
\end{figure}

Figure \ref{fig:prof_perp} shows perpendicular profiles across the current layers at the moment of saturation of coalescence instability, compared for simulations in the `2Da' series with initial plasma temperatures spanning two orders of magnitude.
We show the profiles of three parameters: number density $n$, mean particle energy (`temperature') $\left<\gamma\right>$, and non-ideal electric field scalar $\bm{E}\cdot\bm{B}$.
Consistent with the findings of \cite{nalewajko_kinetic_2016}, the profiles of $\bm{E}\cdot\bm{B}$ are much thicker than the profiles of density or temperature.
Here, we find that radiative cooling has no effect on the profiles of $\bm{E}\cdot\bm{B}$.
On the other hand, its effect on the temperature profile is seen for $\Theta > 10^6$.
In the cases of very strong radiative cooling, the temperature profiles are flattened, approaching the thickness scale of the $\bm{E}\cdot\bm{B}$.
The profiles of plasma density are not consistent between the different cooling rates (i.e., the density peak values do not vary systematically with $\Theta_{\rm e}$), although they seem to preserve the same thickness scale.

\begin{figure}
  \centering
  \includegraphics[width=0.495\textwidth]{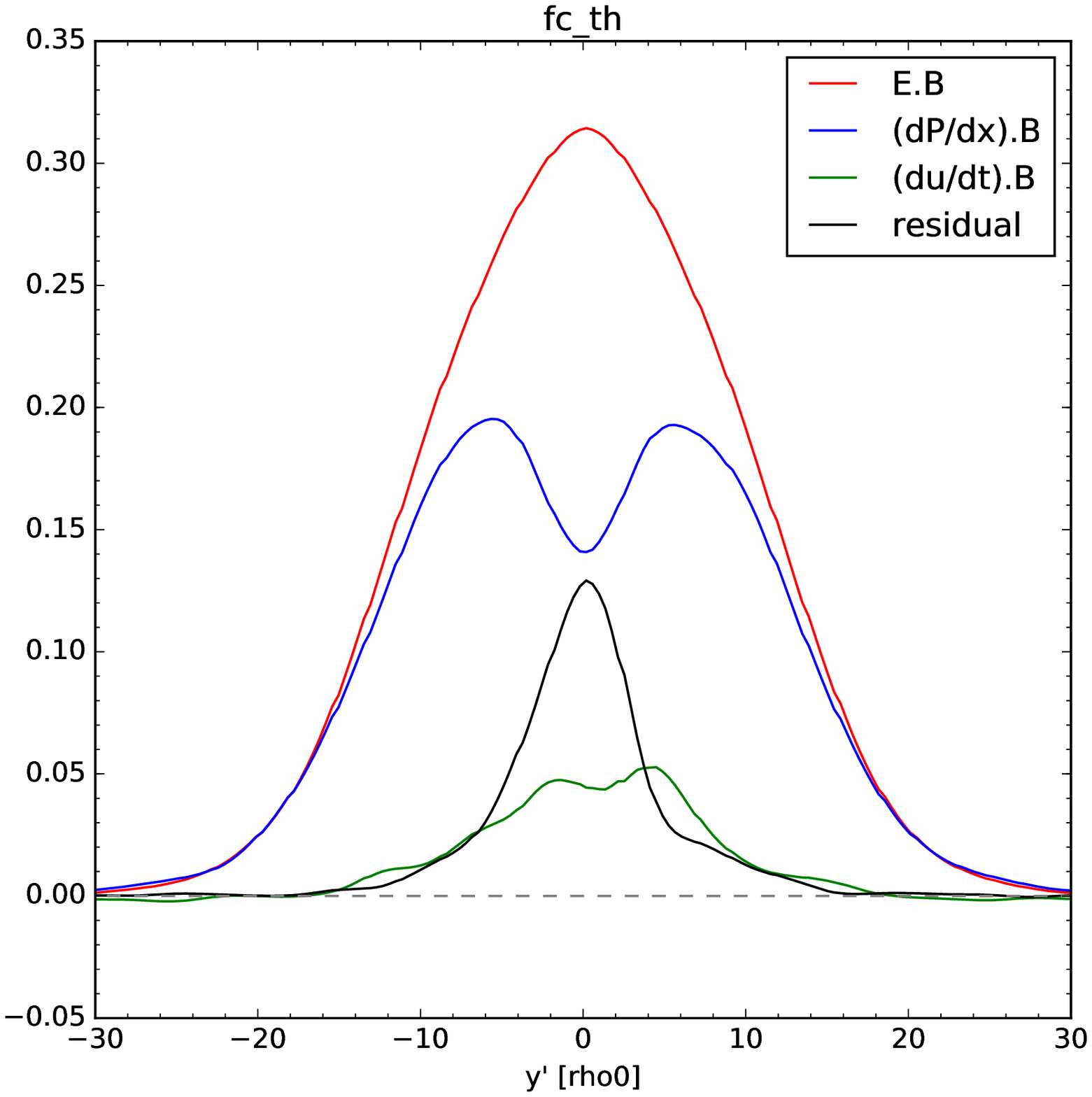}
  \includegraphics[width=0.495\textwidth]{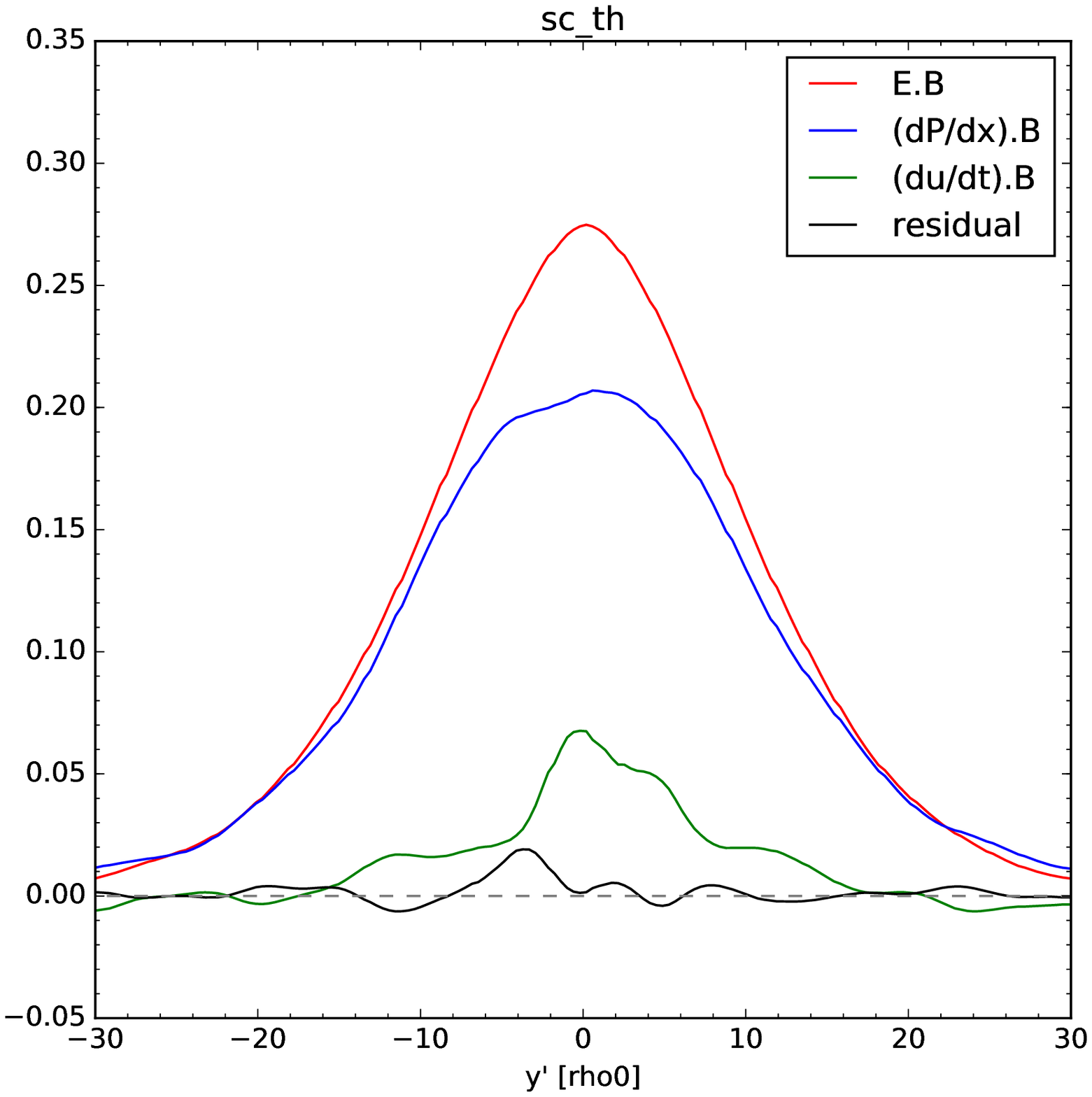}
  \caption{Decomposition of the perpendicular profile of $\bm{E}\cdot\bm{B}$ according to the Vlasov momentum equation (Eq. \ref{eq:vlasov_momentum}) compared for two simulations of the `2Db' series: `fc\_th' (left panel) and `sc\_th' (right panel).}
  \label{fig:prof_vlasov}
\end{figure}

Figure \ref{fig:prof_vlasov} shows the decomposition of the perpendicular profile of $\bm{E}\cdot\bm{B}$ into components of so-called Vlasov momentum equation. We consider a phase-space distribution of particles (either electrons or positrons) $f(\bm{x},\bm{p})$ that satisfies the Vlasov equation ${\rm d}f/{\rm d}t = \partial f/\partial t + v^i(\partial f/\partial x^i) + F^i(\partial f/\partial p^i) = 0$, where $F^i$ is any force acting on individual particles (in this case, it consists of Lorentz force and radiation reaction).
A first-moment equation $\int p^i ({\rm d}f/{\rm dt}){\rm d}^3p = 0$ in scalar product with magnetic field leads to the following form:
\begin{equation}
\label{eq:vlasov_momentum}
\left(E^i - \frac{\partial u^i}{\partial t} - \frac{\partial P^{ij}}{\partial x^j} - Q_{\rm rad}^i\right)B^i = 0
\end{equation}
Here, $\bm{u} = n\left<\bm{p}\right>$ is the momentum density, $P^{ij} = n\left<p^iv^j\right>$ is the pressure tensor, and $\bm{Q}_{\rm rad}$ is a vector related to the radiation reaction force (detailed derivation will be presented elsewhere).
It has been understood previously that gradients of the off-diagonal terms of the pressure tensor are important in supporting the non-ideal electric field in collisionless pair-plasma reconnection \citep{Bessho2007}.
Figure \ref{fig:prof_vlasov} shows that gradients of the pressure tensor balance $\bm{E}\cdot\bm{B}$ in the outer regions of the current layer. 
However, in the central regions of the current layer in the fast-cooling `fc\_th' case, there is a gap between the electric-field and pressure terms, that is only partially filled with the momentum density term.
The residual $(E^i - \partial_tu^i - \partial_jP^{ij})B^i$ appears on the thickness scale of the temperature profiles (cf. Figure \ref{fig:prof_perp}). Since this residual is very small in the slow-cooling `sc\_th' case, we associate it with radiation reaction.

\section{Discussion}
\label{sec_disc}

{In this work, we consider an IC process with the soft photons represented as a fixed uniform isotropic monoenergetic radiation field, corresponding to the ERC scenario characteristic for the luminous FSRQ-type blazars (Appendix \ref{sec_blazars}).
We note that in order to include the SSC mechanism, we would need to calculate the production and propagation of multi-directional synchrotron radiation field, which would require much more complicated and expensive algorithms.
Therefore, we focus here on a very specific application to the broad-band emission from FSRQs.

\subsection{Compton dominance}

In all of our simulations, we have set the energy density of soft radiation fields at the level of mean magnetic energy density $U_{\rm ext} = \left<U_{\rm B}\right>$. The idea was to make radiative cooling due to the synchrotron and IC processes equally important, at least as long as the IC scattering proceeds in the Thomson regime, and so the resulting luminosities should also be comparable, i.e. the Compton dominance parameter ${\rm CD} = L_{\rm IC}/L_{\rm syn}$ should be close to unity by design. In blazars, the Compton dominance is generally observed to range between 0.1 and 100 (Appendix \ref{sec_blazars}).

It is hoped that the observed values of Compton dominance can be used to constrain the physical parameters of blazar jets, especially in the context of luminous FSRQ blazars where gamma rays are produced by Comptonization of external radiation fields. With independent estimates of $U_{\rm ext}$ (from observations of broad emission lines or thermal emission of accretion disks), the value of ${\rm CD}$ provides an independent constrain on magnetic field strength (that determines $L_{\rm syn}$) and hence the jet magnetisation (with total jet power constrained by $L_{\rm IC}$) \citep{Janiak2015}. However, this reasoning relies on the assumption that ${\rm CD} \simeq U_{\rm ext} / U_{\rm B}$, which we are now testing in the case of electrons accelerated in relativistic magnetic reconnection.

Our simulated synchrotron and IC spectra show a range of behaviours leading to departures of Compton dominance from unity in both directions. In the case of IC scattering in the Klein-Nishina regime, ${\rm CD} < 1$ from the beginning due to suppression of the scattering cross section. As the IC spectral peak is dominated by contribution from cold electrons, the evolution of ${\rm CD}$ is driven primarily by evolution of the synchrotron peak. However, with IC scattering in the Thomson regime, the results are more nuanced. In the slow-cooling case (sc\_th), energetic electrons produce a more prominent synchrotron excess, slightly suppressing the value of ${\rm CD}$. On the other hand, in the fast-cooling case (fc\_th), the high-frequency IC excess is more prominent, and the value of ${\rm CD}$ increases to $\sim 10$. We think that the main reason for this is that in the fast-cooling regime, energetic particles are strongly concentrated within current layers where (perpendicular) magnetic fields are much weaker and synchrotron losses are suppressed until the particles are able to exit the current layers. This separation between acceleration and radiation phases has been seen previously in \cite{yuan_kinetic_2016}, where only the synchrotron cooling was considered. However, if IC cooling is efficient (in the Thomson regime), it can easily dominate within the current layers. We should also note that this effect is not seen in our 3D simulation, where cooling efficiency is comparable and IC scattering is more strictly Thomson. We have indication that the acceleration regions are less regular and they may not be able to trap energetic particles as well as in the 2D case.

Variations of Compton dominance are observed even for individual sources when compared at different flux states, the highest values are typically observed during the most spectacular gamma-ray flares \citep{hayashida_rapid_2015}. It has been previously suggested that the highest observed values of Compton dominance could result from localised destruction of magnetic fields by relativistic reconnection. Our results suggest that this could indeed happen (at least in the 2D approximation) if energetic particles are confined to their acceleration regions where magnetic fields are decreased due to reconnection or forced to propagate along the magnetic field lines by parallel electric fields.

\subsection{Variability of lightcurves}

We have simulated synchrotron and IC radiation signals (lightcurves) produced by the same population of energetic electrons and calculated the corresponding Discrete Correlation Functions (DCF). In the fast-cooling regime, we produce short simultaneous (time lags $<0.1L/c$) flares of synchrotron and IC radiation. In the slow-cooling regimes, we obtain quasi-periodic oscillations with the period of $\simeq L/(2c)$, which reflects the periodic nature of the simulated system and the fact that particles crossing the periodic boundary normal to the observer line produce radiation signals separated by a delay of $L/c$. In the fast-cooling 2D cases, the emitting regions can be to some degree isolated from the periodic boundaries. This is more difficult in the 3D case, which features a larger number of magnetic reconnection sites that contribute to the observed signals. Hence, we do not find evidence for systematic time delay between synchrotron and IC signals produced by the same electrons.

The Power Spectral Density (PSD) analysis reveals that in 2D simulations the simulated signals are consistent with power laws with indices between $-1$ and $-2$, as is generally observed in blazars (Appendix \ref{sec_blazars}). However, the 3D simulation produced smoother lightcurves with steep PSD of index $\simeq -3$. This signal could result from averaging comparable contributions from multiple emitting regions located in a relatively small simulation volume. The 3D ABC configuration with $k = 2$ may simply be too regular to produce realistic radiation signals, as no individual reconnection site can dominate the observed signal.
}

\subsection{Effects of radiative losses on development of current layers}

Simulations of magnetic dissipation initialised with the `ABC fields' are particularly valuable for studies of `ab-initio' formation of dynamically evolving current layers. In the previous study \citep{nalewajko_kinetic_2016}, we have found that such current layers develop two separate perpendicular thickness scales: a thinner density scale corresponding to the plasma skin-depth, and a thicker $\bm{E}\cdot\bm{B}$ scale corresponding to the gyroradius of particles heated by the non-ideal electric field within the layer.

Our new results shown in Figure \ref{fig:prof_perp} indicate that these separate thickness scales persist even under severe radiative cooling.
We have also found evidence that the temperature thickness scale can be modified by strong radiative cooling, and hence it is not fundamentally related to the density thickness scale.
Dynamical effects of strong radiative cooling are also found in the decomposition of the $\bm{E}\cdot\bm{B}$ profile across the current layer (Figure \ref{fig:prof_vlasov}), using the Vlasov momentum equation that anticipates the contribution of radiation reaction.

Effects of intense radiative cooling on the structure of reconnection current layers were investigated analytically by \cite{uzdensky_magnetic_2011} in the framework of the classical Sweet-Parker model of resistive reconnection.
Their main prediction is that of systematic compression of plasma within the current layer, $A = n/n_0$, which would effectively enhance the reconnection rate like $v_{\rm rec} \propto A^{1/2}$.
Our results obtained for the `2Da' series of simulations and presented in Figure \ref{fig:prof_perp} reveal neither a systematic increase in the plasma density with increasing $\Theta_{\rm e}$ or decreasing $l_{\rm cool}$, nor a systematic increase in the reconnection rate, an evidence for which would be revealed by the peak levels of the $\bm{E}\cdot\bm{B}$ profiles.

We can formally estimate the value of predicted compression ratio $A$.
Using Equations (45-46) of \cite{uzdensky_magnetic_2011} with our Equation (\ref{eq:lcool}), the following expressions can be found for the volumetric energy conversion rates: $Q_0 = v_{\rm A0}B_0^2/(4\pi L_{\rm layer})$ and $Q_{\rm rad} = \frac{4}{3}\sigma_{\rm T}cu^2U_0n = \Theta_{\rm e}nm_{\rm e}c^3/l_{\rm cool} \simeq cB_0^2/(8\pi l_{\rm cool})$, and hence $A = Q_{\rm rad}/Q_0 \simeq \frac{1}{2}(c/v_{\rm A0})(L_{\rm layer}/l_{\rm cool})$, where $v_{\rm A0} \sim c$ is the Alfv\'{e}n wave velocity in the relativistically magnetised background plasma.
In our 2D simulations initiated with diagonal ABC fields, typical half-length of the saturated current layer is $L_{\rm layer} \sim 300\rho_0$, hence we have at least nominally achieved conditions for $A > 1$ only for 2 simulations in the `2Da' series for $\Theta_{\rm e} \ge 3\times 10^6$, for which $l_{\rm cool} < 300\rho_0$ (see Table \ref{tab:sims}).
Indeed, these are the two simulations, for which we find modified temperature profiles across the current layers.

\appendix
\section{Synchrotron and IC radiation of blazars}
\label{sec_blazars}

{The most direct application of this work concerns the broad-band non-thermal emission from extragalactic astrophysical sources called blazars.
Here we provide a brief summary of the physical parameters of their emission regions (see \citealt{madejski_gamma-ray_2016} for a recent review).

Blazars are a subclass of active galactic nuclei, which are accreting supermassive black holes (of typical mass $M_{\rm bh} \sim 10^8-10^9 M_\odot$) residing in the centres of certain galaxies. The emission of blazars is strongly dominated by broad-band non-thermal continua forming two main components: (1) the low-energy component extends from radio waves to UV/X-ray and is universally attributed to synchrotron radiation; (2) the high-energy component peaks in the MeV/GeV/TeV gamma-ray band, and is generally attributed to leptonic IC radiation, although hadronic radiative processes are also being proposed \citep{Sikora2009,Boettcher2013}. The bolometric apparent luminosities of blazars range from $10^{43}\;{\rm erg/s}$ to $10^{49}\;{\rm erg/s}$, exceeding the luminosities of their host galaxies (up to $10^{45}\;{\rm erg/s}$).
The luminosity ratio $L_{\rm IC}/L_{\rm syn}$ of the IC and synchrotron spectral components is known as the Compton dominance.
The observed values of Compton dominance range from 0.1-1 in the case of low-luminosity (BL Lac type) blazars up to 100 in the case of high-luminosity (Flat Spectrum Radio Quasars; FSRQs) blazars \citep{Finke2013,Nalewajko_Gupta2017}.
At the same time, the non-thermal continua of blazars show persistent chaotic variability over time scales ranging from decades to minutes.
The Power Spectral Densities (PSD) of actual blazar lightcurves are consistent with featureless power laws with indices between $-1$ and $-2$ \citep{Abdo2010_varstats,Goyal2017}.
Variability of the high-energy IC component (gamma-rays) is in general well correlated with variability of the low-energy synchrotron component (X-ray/optical), especially if one compares signals produced by electrons in the similar energy range.
In some cases, the correlations are strictly simultaneous, as confirmed by the Discrete Correlation Function (DCF) analysis \citep{Wehrle2012_3c454}. 
For the above reasons, this non-thermal emission is attributed to relativistic jets streaming from the supermassive black holes with typical bulk Lorentz factors $\Gamma_{\rm j} \sim 5-40$ and (by chance) pointing towards the observer. Radiation produced isotropically within relativistic jets is very strongly beamed (the apparent bolometric luminosity scales roughly like $L_{\rm app} \sim \Gamma_{\rm j}^4 L_{\rm int}'$), at the same time the variability time scale is shortened like $T_{\rm app} \sim T_{\rm int}/\Gamma_{\rm j}$.

The extreme broadness of non-thermal blazar spectra requires a population of ultra-relativistic electrons reaching typical isotropic Lorentz factors (in the rest frame of bulk jet flow) of $\gamma_e \sim 10^3-10^5$. A soft radiation field of co-moving photon energy $E_{\rm soft}'$ will be upscattered to a gamma-ray photon of typical energy $E_\gamma' \sim \gamma_e^2E_{\rm soft}'$. This means that the observed gamma rays are typically produced in a single scattering of a UV/optical/IR soft photon. The soft photons can be local synchrotron photons -- such IC process is called Synchrotron Self-Compton (SSC) and is thought to dominate in the low-luminosity (BL Lac type) blazars. Alternatively, the soft photons may originate outside the relativistic jet -- such IC process is called External Radiation Compton (ERC) and is thought to dominate in the high-luminosity (FSRQs) blazars. Luminous blazars are known to have a very rich radiative environment including direct thermal (UV/optical) emission of the accretion disk (quasar), broad emission lines (e.g. Ly$\alpha$ at $\simeq 10\;{\rm eV}$), and thermal IR emission of the dusty torus ($\sim 0.3\;{\rm eV}$).

It is typically inferred from detailed modelling of observed FSRQ spectra that they should be produced at distance scales in the range $r \sim 0.1-10\;{\rm pc}$ from the black hole \citep{nalewajko_constraining_2014}. The total inferred power of the underlying relativistic jet can reach Eddington luminosity $P_{\rm j} \sim L_{\rm Edd} \sim 10^{46}\;{\rm erg/s}$. We can adopt a conical geometry with opening angle $\Theta_j \sim 0.3/\Gamma_j$. The actual value of jet hot magnetisation $\sigma_{\rm hot,j} = B_j'^2/(4\pi w_j)$ (where $B_j'$ is the co-moving jet magnetic field strength and $w_j$ is the co-moving relativistic enthalpy density) in the dissipation/emission regions is uncertain. In fact, it is still being debated whether dissipation should be provided by shock waves at low magnetizations or by magnetic reconnection at high magnetizations, in any case one needs to distinguish between background/upstream and processed/downstream (see a discussion in \citealt{sironi_relativistic_2015}). We can assume that $\sigma_{\rm hot,j} \sim 1$ and estimate the co-moving magnetic field strength from the magnetic jet power $L_B = cR_j^2\Gamma_j^2B_j'^2/8 \sim L_{\rm Edd}/4$, with jet radius $R_j = \Theta_jr$, hence $B_j' \sim (2L_{\rm Edd}/c)^{1/2}/(0.3r) \sim 1\;{\rm G} (r/1\;{\rm pc})^{-1}$. With such magnetic field strength, the characteristic electron gyroradius is $\rho_e = 1.7\times 10^7(r/1\;{\rm pc})(\gamma_e/10^4)\;{\rm cm}$, about 10 orders of magnitude smaller than the characteristic jet radius $R_j \sim 10^{17}(r/1\;{\rm pc})\;{\rm cm}$.

While we can estimate the relativistic specific enthalpy of the jet as $w_j = B_j'^2/(4\pi\sigma_{\rm hot,j})$, the number density of electrons depends on the poorly understood composition of jet plasma. Relativistic jets can be composed of protons (ions) and ions and electrons with significant addition of electron-positron pairs \citep{Sikora2009}. If the energetic coupling between protons and leptons is strong, they can both participate in non-thermal particle acceleration and share the dissipated magnetic/kinetic energy.
We can effectively parametrize this unknown microphysics with $g_p = \left<\gamma_e\right> + \left<\gamma_p\right>(n_pm_p/n_em_e)$, and then find the electron number density from relation $w_j \simeq 4g_pn_em_ec^2$.
The electron number density scales like $n_e = B_j'^2/(16\pi\sigma_{\rm hot,j}g_pm_ec^2) \sim 5\times 10^4\;{\rm cm^{-3}}g_p^{-1}(r/1\;{\rm pc})^{-2}$.
The Thomson optical depth across the jet is $\tau_{\rm T,j} = \sigma_{\rm T}n_eR_j' \sim 0.003g_p^{-1}(r/1\;{\rm pc})^{-1}$.
Given that most likely $g_p \gg 1$, relativistic jets of blazars are optically thin at parsec scales.
}

\vskip 1em
\paragraph{Acknowledgements:}
{We thank the anonymous reviewers for their helpful comments and suggestions.}
{\tt Zeltron} is an open explicit parallelised particle-in-cell numerical code created by Beno{\^i}t Cerutti and co-developed by Gregory Werner at the University of Colorado Boulder (\url{http://benoit.cerutti.free.fr/Zeltron/}).
These results are based on numerical simulations performed at supercomputers: \emph{Mira} at Argonne National Laboratory, USA (2016 INCITE allocation; PI: D. Uzdensky) and \emph{Prometheus} at Cyfronet AGH, Poland.
This work was supported by the Polish National Science Centre grant 2015/18/E/ST9/00580. YY gratefully acknowledges the support through a Lyman Spitzer, Jr. Postdoctoral Fellowship from the Department of Astrophysical Sciences, Princeton University.

\bibliographystyle{jpp}

\begin{thebibliography}{40}
\expandafter\ifx\csname natexlab\endcsname\relax\def\natexlab#1{#1}\fi
\def\au#1{#1} \def\ed#1{#1} \def\yr#1{#1}\def\at#1{#1}\def\jt#1{\textit{#1}}
  \def\bt#1{#1}\def\bvol#1{\textbf{#1}} \def\vol#1{#1} \def\pg#1{#1}
  \def\publ#1{#1}\def\arxiv#1{#1}\def\org#1{#1}\def\st#1{\textit{#1}}

\bibitem[{Abdo} \& et~al.(2010)]{Abdo2010_varstats}
{\sc \au{{Abdo}, A.~A.} \& \au{et~al.}} \yr{2010}  \at{{Gamma-ray Light Curves
  and Variability of Bright Fermi-detected Blazars}}.  \jt{The Astrophysical
  Journal}  \bvol{722},  \pg{520--542}.

\bibitem[Ackermann \& et~al.(2016)]{ackermann_minute-timescale_2016}
{\sc \au{Ackermann, M.} \& \au{et~al.}} \yr{2016}  \at{Minute-timescale
  {\textgreater}100 {MeV} γ-{Ray} {Variability} during the {Giant} {Outburst}
  of {Quasar} 3c 279 {Observed} by {Fermi}-{LAT} in 2015 {June}}.  \jt{The
  Astrophysical Journal Letters}  \bvol{824},  \pg{L20}.

\bibitem[{Ajello} \& et~al.(2015)]{Ajello2015}
{\sc \au{{Ajello}, M.} \& \au{et~al.}} \yr{2015}  \at{{The Origin of the
  Extragalactic Gamma-Ray Background and Implications for Dark Matter
  Annihilation}}.  \jt{The Astrophysical Journal Letters}  \bvol{800},
  \pg{L27}.

\bibitem[{Bessho} \& {Bhattacharjee}(2007)]{Bessho2007}
{\sc \au{{Bessho}, N.} \& \au{{Bhattacharjee}, A.}} \yr{2007}  \at{{Fast
  collisionless reconnection in electron-positron plasmas}}.  \jt{Physics of
  Plasmas}  \bvol{14}~(5),  \pg{056503}.

\bibitem[Blandford {\em et~al.\/}(2017)Blandford, Yuan, Hoshino \&
  Sironi]{blandford_magnetoluminescence_2017}
{\sc \au{Blandford, R.}, \au{Yuan, Y.}, \au{Hoshino, M.} \& \au{Sironi, L.}}
  \yr{2017}  \at{Magnetoluminescence}.  \jt{Space Science Reviews}  \bvol{207},
   \pg{291--317}.

\bibitem[{Blumenthal} \& {Gould}(1970)]{Blumenthal_Gould1970}
{\sc \au{{Blumenthal}, G.~R.} \& \au{{Gould}, R.~J.}} \yr{1970}
  \at{Bremsstrahlung, synchrotron radiation, and compton scattering of
  high-energy electrons traversing dilute gases}.  \jt{Review of Modern
  Physics}  \bvol{42},  \pg{237--271}.

\bibitem[{B{\"o}ttcher} {\em et~al.\/}(2013){B{\"o}ttcher}, {Reimer}, {Sweeney}
  \& {Prakash}]{Boettcher2013}
{\sc \au{{B{\"o}ttcher}, M.}, \au{{Reimer}, A.}, \au{{Sweeney}, K.} \&
  \au{{Prakash}, A.}} \yr{2013}  \at{{Leptonic and Hadronic Modeling of
  Fermi-detected Blazars}}.  \jt{The Astrophysical Journal}  \bvol{768},
  \pg{54}.

\bibitem[Buehler \& {et~al.}(2012)]{buehler_gamma-ray_2012}
{\sc \au{Buehler, R.} \& \au{{et~al.}}} \yr{2012}  \at{Gamma-{Ray} {Activity}
  in the {Crab} {Nebula}: {The} {Exceptional} {Flare} of 2011 {April}}.
  \jt{The Astrophysical Journal}  \bvol{749},  \pg{26}.

\bibitem[Cerutti {\em et~al.\/}(2016)Cerutti, Philippov \&
  Spitkovsky]{cerutti_modelling_2016}
{\sc \au{Cerutti, B.}, \au{Philippov, A.~A.} \& \au{Spitkovsky, A.}} \yr{2016}
  \at{Modelling high-energy pulsar light curves from first principles}.
  \jt{Monthly Notices of the Royal Astronomical Society}  \bvol{457},
  \pg{2401--2414}.

\bibitem[Cerutti {\em et~al.\/}(2013)Cerutti, Werner, Uzdensky \&
  Begelman]{cerutti_simulations_2013}
{\sc \au{Cerutti, B.}, \au{Werner, G.~R.}, \au{Uzdensky, D.~A.} \&
  \au{Begelman, M.~C.}} \yr{2013}  \at{Simulations of {Particle} {Acceleration}
  beyond the {Classical} {Synchrotron} {Burnoff} {Limit} in {Magnetic}
  {Reconnection}: {An} {Explanation} of the {Crab} {Flares}}.  \jt{The
  Astrophysical Journal}  \bvol{770},  \pg{147}.

\bibitem[Cerutti {\em et~al.\/}(2014)Cerutti, Werner, Uzdensky \&
  Begelman]{cerutti_three-dimensional_2014}
{\sc \au{Cerutti, B.}, \au{Werner, G.~R.}, \au{Uzdensky, D.~A.} \&
  \au{Begelman, M.~C.}} \yr{2014}  \at{Three-dimensional {Relativistic} {Pair}
  {Plasma} {Reconnection} with {Radiative} {Feedback} in the {Crab} {Nebula}}.
  \jt{The Astrophysical Journal}  \bvol{782},  \pg{104}.

\bibitem[{Dombre} {\em et~al.\/}(1986){Dombre}, {Frisch}, {Henon}, {Greene} \&
  {Soward}]{Dombre1986}
{\sc \au{{Dombre}, T.}, \au{{Frisch}, U.}, \au{{Henon}, M.}, \au{{Greene},
  J.~M.} \& \au{{Soward}, A.~M.}} \yr{1986}  \at{{Chaotic streamlines in the
  ABC flows}}.  \jt{Journal of Fluid Mechanics}  \bvol{167},  \pg{353--391}.

\bibitem[East {\em et~al.\/}(2015)East, Zrake, Yuan \&
  Blandford]{east_spontaneous_2015}
{\sc \au{East, W.~E.}, \au{Zrake, J.}, \au{Yuan, Y.} \& \au{Blandford, R.~D.}}
  \yr{2015}  \at{Spontaneous {Decay} of {Periodic} {Magnetostatic}
  {Equilibria}}.  \jt{Physical Review Letters}  \bvol{115},  \pg{095002}.

\bibitem[{Edelson} \& {Krolik}(1988)]{EdelsonKrolik1988}
{\sc \au{{Edelson}, R.~A.} \& \au{{Krolik}, J.~H.}} \yr{1988}  \at{{The
  discrete correlation function - A new method for analyzing unevenly sampled
  variability data}}.  \jt{The Astrophysical Journal}  \bvol{333},
  \pg{646--659}.

\bibitem[{Finke}(2013)]{Finke2013}
{\sc \au{{Finke}, J.~D.}} \yr{2013}  \at{{Compton Dominance and the Blazar
  Sequence}}.  \jt{The Astrophysical Journal}  \bvol{763},  \pg{134}.

\bibitem[{Goyal} {\em et~al.\/}(2017){Goyal}, {Stawarz}, {Ostrowski},
  {Larionov}, {Gopal-Krishna}, {Wiita}, {Joshi}, {Soida} \& {Agudo}]{Goyal2017}
{\sc \au{{Goyal}, A.}, \au{{Stawarz}, {\L}.}, \au{{Ostrowski}, M.},
  \au{{Larionov}, V.}, \au{{Gopal-Krishna}}, \au{{Wiita}, P.~J.}, \au{{Joshi},
  S.}, \au{{Soida}, M.} \& \au{{Agudo}, I.}} \yr{2017}  \at{{Multiwavelength
  Variability Study of the Classical BL Lac Object PKS 0735+178 on Timescales
  Ranging from Decades to Minutes}}.  \jt{The Astrophysical Journal}
  \bvol{837},  \pg{127}.

\bibitem[Guo {\em et~al.\/}(2014)Guo, Li, Daughton \& Liu]{guo_formation_2014}
{\sc \au{Guo, F.}, \au{Li, H.}, \au{Daughton, W.} \& \au{Liu, Y.-H.}} \yr{2014}
   \at{Formation of {Hard} {Power} {Laws} in the {Energetic} {Particle}
  {Spectra} {Resulting} from {Relativistic} {Magnetic} {Reconnection}}.
  \jt{Physical Review Letters}  \bvol{113},  \pg{155005}.

\bibitem[Hayashida \& et~al.(2015)]{hayashida_rapid_2015}
{\sc \au{Hayashida, M.} \& \au{et~al.}} \yr{2015}  \at{Rapid {Variability} of
  {Blazar} 3c 279 during {Flaring} {States} in 2013-2014 with {Joint}
  {Fermi}-{LAT}, {NuSTAR}, {Swift}, and {Ground}-{Based} {Multiwavelength}
  {Observations}}.  \jt{The Astrophysical Journal}  \bvol{807},  \pg{79}.

\bibitem[{Hededal} \& {Nordlund}(2005)]{Hededal2005}
{\sc \au{{Hededal}, C.~B.} \& \au{{Nordlund}, {\AA}.}} \yr{2005}
  \at{{Gamma-Ray Burst Synthetic Spectra from Collisionless Shock PIC
  Simulations}}.  \jt{ArXiv Astrophysics e-prints} ,  \arxiv{arXiv:
  astro-ph/0511662}.

\bibitem[{Janiak} {\em et~al.\/}(2015){Janiak}, {Sikora} \&
  {Moderski}]{Janiak2015}
{\sc \au{{Janiak}, M.}, \au{{Sikora}, M.} \& \au{{Moderski}, R.}} \yr{2015}
  \at{{Magnetization of jets in luminous blazars}}.  \jt{Monthly Notices of the
  Royal Astronomical Society}  \bvol{449},  \pg{431--439}.

\bibitem[{Jones}(1968)]{Jones1968}
{\sc \au{{Jones}, F.~C.}} \yr{1968}  \at{{Calculated Spectrum of
  Inverse-Compton-Scattered Photons}}.  \jt{Physical Review}  \bvol{167},
  \pg{1159--1169}.

\bibitem[Kagan {\em et~al.\/}(2016)Kagan, Nakar \& Piran]{kagan_beaming_2016}
{\sc \au{Kagan, D.}, \au{Nakar, E.} \& \au{Piran, T.}} \yr{2016}  \at{Beaming
  of {Particles} and {Synchrotron} {Radiation} in {Relativistic} {Magnetic}
  {Reconnection}}.  \jt{The Astrophysical Journal}  \bvol{826},  \pg{221}.

\bibitem[{Lyutikov} {\em et~al.\/}(2018){Lyutikov}, {Komissarov}, {Sironi} \&
  {Porth}]{Lyutikov2018}
{\sc \au{{Lyutikov}, M.}, \au{{Komissarov}, S.}, \au{{Sironi}, L.} \&
  \au{{Porth}, O.}} \yr{2018}  \at{{Particle acceleration in explosive
  relativistic reconnection events and Crab Nebula gamma-ray flares}}.
  \jt{ArXiv e-prints} ,  \arxiv{arXiv: 1804.10291}.

\bibitem[{Lyutikov} {\em et~al.\/}(2017){Lyutikov}, {Sironi}, {Komissarov} \&
  {Porth}]{lyutikov_particle_2017}
{\sc \au{{Lyutikov}, M.}, \au{{Sironi}, L.}, \au{{Komissarov}, S.~S.} \&
  \au{{Porth}, O.}} \yr{2017}  \at{{Particle acceleration in relativistic
  magnetic flux-merging events}}.  \jt{Journal of Plasma Physics}
  \bvol{83}~(6),  \pg{635830602}.

\bibitem[Madejski \& Sikora(2016)]{madejski_gamma-ray_2016}
{\sc \au{Madejski, G.~G.} \& \au{Sikora, M.}} \yr{2016}  \at{Gamma-{Ray}
  {Observations} of {Active} {Galactic} {Nuclei}}.  \jt{Annual Review of
  Astronomy and Astrophysics}  \bvol{54},  \pg{725--760}.

\bibitem[Mizuno {\em et~al.\/}(2009)Mizuno, Lyubarsky, Nishikawa \&
  Hardee]{mizuno_three-dimensional_2009}
{\sc \au{Mizuno, Y.}, \au{Lyubarsky, Y.}, \au{Nishikawa, K.-I.} \& \au{Hardee,
  P.~E.}} \yr{2009}  \at{Three-{Dimensional} {Relativistic}
  {Magnetohydrodynamic} {Simulations} of {Current}-{Driven} {Instability}. {I}.
  {Instability} of a {Static} {Column}}.  \jt{The Astrophysical Journal}
  \bvol{700},  \pg{684--693}.

\bibitem[Nalewajko {\em et~al.\/}(2014)Nalewajko, Begelman \&
  Sikora]{nalewajko_constraining_2014}
{\sc \au{Nalewajko, K.}, \au{Begelman, M.~C.} \& \au{Sikora, M.}} \yr{2014}
  \at{Constraining the {Location} of {Gamma}-{Ray} {Flares} in {Luminous}
  {Blazars}}.  \jt{The Astrophysical Journal}  \bvol{789},  \pg{161}.

\bibitem[{Nalewajko} \& {Gupta}(2017)]{Nalewajko_Gupta2017}
{\sc \au{{Nalewajko}, K.} \& \au{{Gupta}, M.}} \yr{2017}  \at{{The sequence of
  Compton dominance in blazars based on data from WISE and Fermi-LAT}}.
  \jt{Astronomy \& Astrophysics}  \bvol{606},  \pg{A44}.

\bibitem[Nalewajko {\em et~al.\/}(2016)Nalewajko, Zrake, Yuan, East \&
  Blandford]{nalewajko_kinetic_2016}
{\sc \au{Nalewajko, K.}, \au{Zrake, J.}, \au{Yuan, Y.}, \au{East, W.~E.} \&
  \au{Blandford, R.~D.}} \yr{2016}  \at{Kinetic {Simulations} of the
  {Lowest}-order {Unstable} {Mode} of {Relativistic} {Magnetostatic}
  {Equilibria}}.  \jt{The Astrophysical Journal}  \bvol{826},  \pg{115}.

\bibitem[{Nishikawa}(2016)]{Nishikawa2016}
{\sc \au{{Nishikawa}, K.-I. e.~a.}} \yr{2016}  \at{{Evolution of Global
  Relativistic Jets: Collimations and Expansion with kKHI and the Weibel
  Instability}}.  \jt{The Astrophysical Journal}  \bvol{820},  \pg{94}.

\bibitem[{Philippov} {\em et~al.\/}(2015){Philippov}, {Spitkovsky} \&
  {Cerutti}]{Philippov2015}
{\sc \au{{Philippov}, A.~A.}, \au{{Spitkovsky}, A.} \& \au{{Cerutti}, B.}}
  \yr{2015}  \at{{Ab Initio Pulsar Magnetosphere: Three-dimensional
  Particle-in-cell Simulations of Oblique Pulsars}}.  \jt{The Astrophysical
  Journal}  \bvol{801},  \pg{L19}.

\bibitem[{Sikora} {\em et~al.\/}(2009){Sikora}, {Stawarz}, {Moderski},
  {Nalewajko} \& {Madejski}]{Sikora2009}
{\sc \au{{Sikora}, M.}, \au{{Stawarz}, {\L}.}, \au{{Moderski}, R.},
  \au{{Nalewajko}, K.} \& \au{{Madejski}, G.~M.}} \yr{2009}  \at{Constraining
  emission models of luminous blazar sources}.  \jt{The Astrophysical Journal}
  \bvol{704},  \pg{38--50}.

\bibitem[Sironi {\em et~al.\/}(2016)Sironi, Giannios \&
  Petropoulou]{sironi_plasmoids_2016}
{\sc \au{Sironi, L.}, \au{Giannios, D.} \& \au{Petropoulou, M.}} \yr{2016}
  \at{Plasmoids in relativistic reconnection, from birth to adulthood: first
  they grow, then they go}.  \jt{Monthly Notices of the Royal Astronomical
  Society}  \bvol{462},  \pg{48--74}.

\bibitem[Sironi {\em et~al.\/}(2015)Sironi, Petropoulou \&
  Giannios]{sironi_relativistic_2015}
{\sc \au{Sironi, L.}, \au{Petropoulou, M.} \& \au{Giannios, D.}} \yr{2015}
  \at{Relativistic jets shine through shocks or magnetic reconnection?}
  \jt{Monthly Notices of the Royal Astronomical Society}  \bvol{450},
  \pg{183--191}.

\bibitem[Sironi \& Spitkovsky(2014)]{sironi_relativistic_2014}
{\sc \au{Sironi, L.} \& \au{Spitkovsky, A.}} \yr{2014}  \at{Relativistic
  {Reconnection}: {An} {Efficient} {Source} of {Non}-thermal {Particles}}.
  \jt{The Astrophysical Journal Letters}  \bvol{783},  \pg{L21}.

\bibitem[Uzdensky \& McKinney(2011)]{uzdensky_magnetic_2011}
{\sc \au{Uzdensky, D.~A.} \& \au{McKinney, J.~C.}} \yr{2011}  \at{Magnetic
  reconnection with radiative cooling. {I}. {Optically} thin regime}.
  \jt{Physics of Plasmas}  \bvol{18},  \pg{042105--042105}.

\bibitem[{Wehrle} {\em et~al.\/}(2012){Wehrle}, {Marscher}, {Jorstad},
  {Gurwell}, {Joshi}, {MacDonald}, {Williamson}, {Agudo} \&
  {Grupe}]{Wehrle2012_3c454}
{\sc \au{{Wehrle}, A.~E.}, \au{{Marscher}, A.~P.}, \au{{Jorstad}, S.~G.},
  \au{{Gurwell}, M.~A.}, \au{{Joshi}, M.}, \au{{MacDonald}, N.~R.},
  \au{{Williamson}, K.~E.}, \au{{Agudo}, I.} \& \au{{Grupe}, D.}} \yr{2012}
  \at{{Multiwavelength Variations of 3C 454.3 during the 2010 November to 2011
  January Outburst}}.  \jt{The Astrophysical Journal}  \bvol{758},  \pg{72}.

\bibitem[{Werner} {\em et~al.\/}(2018){Werner}, {Philippov} \&
  {Uzdensky}]{Werner2018_IC}
{\sc \au{{Werner}, G.~R.}, \au{{Philippov}, A.~A.} \& \au{{Uzdensky}, D.~A.}}
  \yr{2018}  \at{{Particle acceleration in relativistic magnetic reconnection
  with strong inverse-Compton cooling in pair plasmas}}.  \jt{ArXiv e-prints} ,
   \arxiv{arXiv: 1805.01910}.

\bibitem[Werner {\em et~al.\/}(2016)Werner, Uzdensky, Cerutti, Nalewajko \&
  Begelman]{werner_extent_2016}
{\sc \au{Werner, G.~R.}, \au{Uzdensky, D.~A.}, \au{Cerutti, B.}, \au{Nalewajko,
  K.} \& \au{Begelman, M.~C.}} \yr{2016}  \at{The {Extent} of {Power}-law
  {Energy} {Spectra} in {Collisionless} {Relativistic} {Magnetic}
  {Reconnection} in {Pair} {Plasmas}}.  \jt{The Astrophysical Journal Letters}
  \bvol{816},  \pg{L8}.

\bibitem[Yuan {\em et~al.\/}(2016)Yuan, Nalewajko, Zrake, East \&
  Blandford]{yuan_kinetic_2016}
{\sc \au{Yuan, Y.}, \au{Nalewajko, K.}, \au{Zrake, J.}, \au{East, W.~E.} \&
  \au{Blandford, R.~D.}} \yr{2016}  \at{Kinetic {Study} of
  {Radiation}-reaction-limited {Particle} {Acceleration} {During} the
  {Relaxation} of {Unstable} {Force}-free {Equilibria}}.  \jt{The Astrophysical
  Journal}  \bvol{828},  \pg{92}.

\end{thebibliography}

\end{document}